  \input lanlmac.tex
\overfullrule=0pt
\input epsf.tex
\figno=0

\def\fig#1#2#3{
\par\begingroup\parindent=0pt\leftskip=1cm\rightskip=1cm\parindent=0pt
\baselineskip=11pt
\global\advance\figno by 1
\midinsert
\epsfxsize=#3
\centerline{\epsfbox{#2}}
\vskip 12pt
{\bf Fig. \the\figno:} #1\par
\endinsert\endgroup\par
}

\def\figd#1#2#3#4#5#6{
\par\begingroup\parindent=0pt\leftskip=1cm\rightskip=1cm\parindent=0pt
\baselineskip=11pt
\global\advance\figno by 1
\midinsert
\centerline{ 
a)\epsfxsize=#3 \epsfbox{#2} \hskip #4 
b) \epsfxsize=#6 \epsfbox{#5}}
\vskip 12pt
{\bf Fig. \the\figno:} #1\par
\endinsert\endgroup\par
}

\def\figlabel#1{\xdef#1{\the\figno}}
\def\encadremath#1{\vbox{\hrule\hbox{\vrule\kern8pt\vbox{\kern8pt
\hbox{$\displaystyle #1$}\kern8pt}
\kern8pt\vrule}\hrule}}

\def\figf#1#2#3#4#5#6#7#8{
\par\begingroup\parindent=0pt\leftskip=1cm\rightskip=1cm\parindent=0pt
\baselineskip=11pt
\global\advance\figno by 1
\midinsert
\centerline{\epsfxsize=#6 \epsfbox{#2} \hskip #7 \epsfxsize=#6 \epsfbox{#3}}
\vskip #8
\centerline{\epsfxsize=#6 \epsfbox{#4} \hskip #7 \epsfxsize=#6 \epsfbox{#5}}
\vskip 12pt
{\bf Fig. \the\figno:} #1\par
\endinsert\endgroup\par
}

\def\figlabel#1{\xdef#1{\the\figno}}
\def\encadremath#1{\vbox{\hrule\hbox{\vrule\kern8pt\vbox{\kern8pt
\hbox{$\displaystyle #1$}\kern8pt}
\kern8pt\vrule}\hrule}}

%
\font\zfont = cmss10 
\font\litfont = cmr6

\def\bigone{\hbox{1\kern -.23em {\rm l}}}
\def\ZZ{\hbox{\zfont Z\kern-.4emZ}}
\def\hf{{\litfont {1 \over 2}}}

\def\p{\partial}

\def\b{\beta}

\def\l{\lambda}
\def\m{\mu}

\def\s{\sigma}

\def\z{\zeta }
\def\vp{\varphi}

\def\O{\Omega}
\def\oo{\hat \omega   }

\def\o{\omega }


   \def\z{\zeta }



   \def\CD {{\cal D}}
   
   \def\CF {{\cal F}}

   \def\CR {{\cal R}}

   \def\CW {{\cal W}}

   \def\CZ {{\cal Z}}

   \def\R{\relax{\rm I\kern-.18em R}}
   \font\cmss=cmss10 \font\cmsss=cmss10 at 7pt
   \def\Z{\relax\ifmmode\mathchoice
   {\hbox{\cmss Z\kern-.4em Z}}{\hbox{\cmss Z\kern-.4em Z}}
   {\lower.9pt\hbox{\cmsss Z\kern-.4em Z}}
   {\lower1.2pt\hbox{\cmsss Z\kern-.4em Z}}\else{\cmss Z\kern-.4em
   Z}\fi}
   \def\p{\partial}
   
   \def\11{1\!\! 1}

    \def\hepth{ {\rm hep-th/}}



\def\Pe{\Psi^{_{E}}}

\def\pse{ \psi^{_{E}} }
\def\pee{\psi^{_{E'}}}


\def\xp{z_{_{+}}}
\def\xm{z_{_{-}}}
\def\xpm{z_{_{\pm}}}
\def\xmp{z_{_{\mp}}}

\def\zp{z_{_{+}}}
\def\zm{z_{_{-}}}
\def\zpm{z_{_{\pm}}}

\def\zkpm{z_{\pm k}}

\def\zkpm{z_{_{\pm, k}}}

\def\zipm#1{z_{_{\pm,#1}}}

\def\bop{\omega^{}_{_{+}}}
\def\bom{\omega^{}_{_{-}}}
\def\bopm{\omega^{}_{_{\pm}}}

\def\tp{t_1}
\def\tm{t_{-1}}
\def\tpm{t_{\pm 1}}
\def\tmp{t_{\mp 1}}


\def\Xp{Z_{+}}
\def\Xm{Z_{-}}
\def\Xpm{Z_{\pm }}

\def\psip{\psi_{_{+}}}
\def\psim{\psi_{_{-}}}
\def\psipm{\psi_{_{\pm}}}

\def\Pep{\Pe_{+}}
\def\Pem{\Pe_{-}}
\def\Pepm{\Pe_{\pm }}

\def\psem{\pse_{_{-}}}
\def\psepm{\pse_{_{\pm }}}
\def\psemp{\pse_{_{\mp }}}
\def\peep{\pee_{_{+}}}

\def\peepm{\pee_{_{\pm }}}

\def\plepm{\psi^{_{E}}_{_{\pm ,<}}}

\def\prepm{\psi^{_{E}}_{_{\pm,> }}}

\def\piep#1{\psi^{_{E}}_{_{+, #1}}}
\def\piem#1{\psi^{_{E}}_{_{-, #1}}}
\def\piepm#1{\psi^{_{E}}_{_{\pm,#1 }}}

\def\pieepm#1{\psi^{_{E'}}_{_{\pm ,#1}}}

\def\Pip#1{\Psi^{_{#1}}_{+}}
\def\Pim#1{\Psi^{_{#1}}_{-}}

\def\Pspm{\Psi_{{\pm}}}

\def\Phipm{\Phi_{\pm}}

\def\Phpm#1{\Phi^{#1}_{\pm}}

\def\PPhpm#1{\Phi'^{#1}_{\pm}}





\def\Tr{\,{\rm Tr}\,}

\def\FT{\log\tau }

\def\o{\omega }\def\z{\zeta }

\def\Zb{{\rm \bf Z}}

\def\dpi{{1\over 2\pi}}

\def\fs{\rm ^{_{Fermi\ sea}}}

\def\xx{ {x} }

\def\gr{D^{(r)}_{IJ}}

\def\KK{K}

\def\tv{{\vec t}}


\def\np#1#2#3{{\it Nucl. Phys.} {\bf B#1} (#2) #3}
\def\pl#1#2#3{{\it Phys. Lett. }{\bf B#1} (#2) #3}
\def\prl#1#2#3{{\it Phys. Rev. Lett.}{\bf #1} (#2) #3}
\def\physrev#1#2#3{{\it Phys. Rev.} {\bf D#1} (#2) #3}
\def\prb#1#2#3{{\it Phys. Rev.} {\bf B#1} (#2) #3}

\def\rmatp#1#2#3{{\it Rev. Math. Phys. }{\bf #1} (#2) #3}
\def\cmp#1#2#3{{\it Comm. Math. Phys.} {\bf #1} (#2) #3}
\def\mpl#1#2#3{{\it Mod. Phys. Lett. }{\bf #1} (#2) #3}
\def\ijmp#1#2#3{{\it Int. J. Mod. Phys.} {\bf #1} (#2) #3}
\def\lmp#1#2#3{{\it Lett. Math. Phys.} {\bf #1} (#2) #3}
\def\tmatp#1#2#3{{\it Theor. Math. Phys.} {\bf #1} (#2) #3}
\def\jhep#1#2#3{{\it JHEP} {\bf #1} (#2) #3}
\def\hepth#1{{\tt hep-th/}#1}

\lref\HKK{ J. Hoppe, V. Kazakov, and I. Kostov,
``Dimensionally reduced SYM$_4$ as solvable matrix quantum
mechanics'', \np{571}{2000}{479},  \hepth{9907058}.}
\lref\NKK{V. Kazakov, I. Kostov and  N.  Nekrasov,
``D-particles, matrix integrals and KP hierachy'',
\np{557}{1999}{413},  \hepth{9810035}.}
\lref\KKK{V. Kazakov, I. Kostov, and D. Kutasov,
``A Matrix Model for the Two Dimensional Black Hole",
\np{622}{2002}{141}, \hepth{0101011}.}
\lref\AK{S. Alexandrov and V. Kazakov,
``Correlators in 2D string theory with vortex condensation'',
\np{610}{2001}{77}, \hepth{0104094}.}
\lref\IK{I. Kostov, ``String Equation for String Theory on a Circle'',
\np{624}{2002}{146}, \hepth{0107247}.}

\lref\KAZMIG{V. Kazakov and A. A. Migdal, \np{311}{1988}{171}.}
\lref\BRKA{E. Brezin, V. Kazakov and Al. Zamolodchikov,
\np{338}{1990}{673}.}
\lref\PARISI{ G. Parisi, \pl{238}{1990}{209, 213}.}
\lref\GRMI{ D. Gross and N. Miljkovic, \pl{238}{1990}{217}.}
\lref\GIZI{P. Ginsparg and J. Zinn-Justin, \pl{240}{1990}{333}.}
\lref\BIPZ{ E. Brezin, C. Itzykson, G. Parisi, and J.-B. Zuber,
\cmp{59}{1978}{35}.}

\lref\GRKL{D. Gross and I. Klebanov, \np{344}{1990}{475};
  \np{354}{1990}{459}.}
\lref\BULKA{D. Boulatov and V. Kazakov,  
``One-Dimensional String Theory with Vortices as Upside-Down
Matrix Oscillator'', \ijmp{8}{1993}{809}, \hepth{0012228}. }

\lref\DVV{R. Dijkgraaf, E. Verlinde, and H. Verlinde,
``String propagation in a black hole geometry'',
\np{371}{1992}{269}.}
\lref\FZZ{ V. Fateev, A. Zamolodchikov, and Al. Zamolodchikov,
{\it unpublished}.}
\lref\AKKII{ S. Yu. Alexandrov, V. A. Kazakov, I. K. Kostov, 
{\it work in progress}. }
\lref\MUKHIVAFA{S. Mukhi and C. Vafa,
``Two dimensional black-hole as a topological coset model of 
c=1 string theory'', \np{407}{1993}{667}, \hepth{9301083}.}
\lref\GOSHALVAFA{D. Ghoshal and C. Vafa,
``c=1 String as the Topological Theory of the Conifold'',
\np{453}{1995}{121}, \hepth{9506122}.}
\lref\KZ{V. A. Kazakov and A. Zeytlin,
``On free energy of 2-d black hole in bosonic string theory'', 
\jhep{0106}{2001{}021}, \hepth{0104138}.}
\lref\TECHNER{J. Teschner, 
``The deformed two-dimensional black hole'',
\pl{458}{1999}{257}, \hepth{9902189}. }
\lref\JAP{T. Fukuda and K. Hosomichi,
``Three-point Functions in Sine-Liouville Theory'',
\jhep{0109}{2001}{003}, \hepth{0105217}. }

\lref\MP{G. Moore and M. Plesser,  
``Classical scattering in 1+1 Dimensional string theory'',
\physrev{46}{1992}{1730}, \hepth{9203060}.} 
\lref\MPR{G. Moore, M. Plesser, and S. Ramgoolam,
``Exact S-matrix for 2D string theory'',
\np{377}{1992}{143}, \hepth{9111035}.}
\lref\MOORE{G. Moore, ``Gravitational phase transitions
  and the sine-Gordon model", \hepth{9203061}.}
\lref\DMP{R. Dijkgraaf, G. Moore, and M.R. Plesser,
``The partition function of 2d string theory'',
\np{394}{1993}{356}, \hepth{9208031}.}

\lref\POLCHINSKI{J. Polchinski, ``What is string theory'',
{\it Lectures presented at the 1994 Les Houches Summer School
``Fluctuating Geometries in Statistical Mechanics and Field Theory''},  
\hepth{9411028}.}
\lref\KLEBANOV{I. Klebanov, {\it Lectures delivered at the ICTP
Spring School on String Theory and Quantum Gravity},
Trieste, April 1991, \hepth{9108019}.}
\lref\JEVICKI{A. Jevicki, Developments in 2D string theory,
\hepth/9309115.}
\lref\HSU{E. Hsu and D. Kutasov, ``The Gravitational Sine-Gordon Model'', 
\np{396}{1993}{693}, \hepth{9212023}.}
\lref\MSS{G. Moore, N. Seiberg, and M. Staudacher,
``From loops to states in 2D quantum gravity'',
\np{362}{1991}{665}. }

\lref\JM{M. Jimbo and T. Miwa, ``Solitons and Infinite Dimensional
Lie Algebras'', {\it Publ. RIMS, Kyoto Univ.} {\bf 19}, No. 3
(1983) 943.}
\lref\Hir{R. Hirota, Direct Method in Soliton Theory {\it Solitons},
Ed. by R.K. Bullogh and R.J. Caudrey, Springer, 1980.}
\lref\UT{K. Ueno and K. Takasaki, ``Toda Lattice Hierarchy":
in `Group representations and systems of differential equations',
{\it Adv. Stud. Pure Math.} {\bf 4} (1984) 1.}
\lref\Takasak{K. Takasaki,
{\it Adv. Stud. Pure Math.} {\bf 4} (1984) 139.}
\lref\MukhiImbimbo{ C. Imbimbo and S. Mukhi,
``The topological matrix model of c=1 String", 
\np{449}{1995}{553}, \hepth{9505127}.}
\lref\NTT{ T.Nakatsu, K.Takasaki, and S.Tsujimaru, 
``Quantum and classical aspects of deformed $c=1$ strings'',
\np{443}{1995}{155}, \hepth{9501038}.}
\lref\Takebe{T. Takebe, ``Toda lattice hierarchy and conservation laws'',
\cmp{129}{1990}{129}.}
\lref\EK{T. Eguchi and H. Kanno,
``Toda lattice hierarchy and the topological description of the 
$c=1$ string theory'', \pl{331}{1994}{330}, hep-th/9404056.}
\lref\Krichever{I. Krichever, {\it Func. Anal. i ego pril.},
{\bf 22:3} (1988) 37  (English translation:
{\it Funct. Anal. Appl.} {\bf 22} (1989) 200);
``The  $\tau$-function of the
universal Witham hierarchy, matrix models and topological field
theories'', {\it Comm. Pure Appl. Math.} {\bf 47} (1992),
\hepth{9205110}.}
\lref\TakTakb{K.~Takasaki and T.~Takebe,
``Quasi-classical limit of Toda hierarchy and W-infinity symmetries'',
\lmp{28}{93}{165}, \hepth{9301070}.}
\lref\orlshu{A. Orlov and E. Shulman, \lmp{12}{1986}{171}.}
\lref\Nakatsu{T. Nakatsu, ``On the string equation at $c=1$'', 
\mpl{A9}{1994}{3313}, \hepth{9407096}.}
\lref\TakSE{K. Takasaki,
``Toda lattice hierarchy and generalized string equations'',
\cmp{181}{1996}{131}, \hepth{9506089}.}
\lref\TakTak{K. Takasaki and T. Takebe, 
``Integrable Hierarchies and Dispersionless Limit'',
\rmatp{7}{1995}{743}, \hepth{9405096}.}

\lref\kkvwz{ I. Kostov, I. Krichever, M. Mineev-Veinstein,
P. Wiegmann, and  A. Zabrodin, ``$\tau$-function
for analytic curves", \hepth{0005259}.}
\lref\Zabrodin{ A. Zabrodin, ``Dispersionless limit of Hirota
equations in some problems of complex analysis'',
\tmatp{129}{2001}{1511}; \tmatp{129}{2001}{239}, {\tt math.CV/0104169}.}
\lref\bmrwz{ A. Boyarsky, A. Marshakov,  O. Ruchhayskiy,
P. Wiegmann, and  A. Zabrodin, ``On associativity equations in
dispersionless integrable hierarchies", 
\pl{515}{2001}{483}, \hepth{0105260}.}
\lref\wz{P. Wiegmann and  A. Zabrodin, ``Conformal maps and
dispersionless integrable hierarchies", \cmp{213}{2000}{523},
\hepth{9909147}.}
\lref\FERTIGone{H.A. Fertig, \prb{36}{1987}{7969}. }
\lref\FERTIGtwo{H.A. Fertig, \prb{387}{1988}{996}. }
\lref\WIEGAGAM{O. Agam, E. Bettelheim, P. Wiegmann, and A. Zabrodin,
``Viscous fingering and a shape of an electronic droplet in 
the Quantum Hall regime'', {\tt cond-mat/0111333}.}
\lref\mwz{ M.~Mineev-Weinstein, P.~B.~Wiegmann, A.~Zabrodin, 
``Integrable Structure of Interface Dynamics'', \prl{84}{2000}{5106},
{\tt nlin.SI/0001007}.}

\lref\WITTENGR{E. Witten, ``Ground Ring of two dimensional string theory'',
\np{373}{1992}{187}, \hepth{9108004}.}
\lref\MuchiImbimbo{C. Imbimbo
and S. Mukhi, ``The topological matrix model of c=1 String",
\hepth{9505127}.}

\rightline{SPHT-T02/055}
\Title{
}
{\vbox{\centerline{Time-dependent backgrounds of 2D string theory   }
\centerline{   }
}}
%
%
\centerline{Sergei Yu. Alexandrov,$^{123}$\footnote{$^{\ast}$}
{alexand@spht.saclay.cea.fr}
Vladimir A. Kazakov$^1$\footnote{$^{\circ}$}{{kazakov@lpt.ens.fr}}
and Ivan K. Kostov$^2$\footnote{$^\bullet$}{{kostov@spht.saclay.cea.fr}}}

\centerline{$^1${\it  Laboratoire de Physique Th\'eorique de l'Ecole
Normale Sup\'erieure,\footnote{$^\#$}{Unit\'e de Recherche du
Centre National de la Recherche Scientifique et de  l'Ecole Normale
Sup\'erieure et \`a l'Universit\'e de Paris-Sud.}}}
\centerline{\it 24 rue Lhomond,  75231 Paris CEDEX, France}
\centerline{$^2${\it Service de Physique Th\'eorique,
CNRS - URA 2306, C.E.A. - Saclay,}}
\centerline{  F-91191 Gif-Sur-Yvette CEDEX, France}
\centerline{$^3$ \it V.A.~Fock Department of
Theoretical Physics, St.~Petersburg
University, Russia}

\bigskip
  \vskip 1cm
\baselineskip8pt{

\baselineskip12pt{
\noindent
We study possible backgrounds of 2D string theory using
its equivalence with a system of fermions in upside-down harmonic
potential. Each background corresponds to a certain profile of the
Fermi sea, which can be considered as a deformation of the hyperbolic
profile characterizing the linear dilaton background. Such a
perturbation is generated by a set of commuting flows, which form a
Toda Lattice integrable structure. The flows are associated with all
possible left and right moving tachyon states, which in the
compactified theory have discrete spectrum. The simplest nontrivial
background describes the Sine-Liouville string theory.  Our methods
can be also applied to the study of 2D droplets of electrons in a
strong magnetic field. }

\Date{May, 2002} 

\baselineskip=14pt plus 2pt minus 2pt

    \newsec{Introduction}

One of the most important problems of the modern string theory is a
search for new nontrivial backgrounds and the study of the underlying
string dynamics. In most of the cases the target space metric of such
backgrounds is curved and often it incorporates the black hole
singularities.  In the superstring theories, the supersymmetry allows
for some interesting nontrivial solutions which are stable and
exact. But the string theory on such backgrounds is usually an
extremely complicated sigma-model, very difficult even to formulate it
explicitly, not to mention studying quantitatively its dynamics.

The two-dimensional bosonic string theory is a rare case of sigma-model
where such a dynamics is integrable, at least for some particular
backgrounds, including the dilatonic black hole background.  A
physically transparent way to study the perturbative (one loop) string
theory around such a background is provided by the CFT approach.
However, once we want to achieve more ambitious quantitative goals,
especially in analyzing higher loops or multipoint correlators, we
have to address ourselves to the matrix model approach to the 2D
string theory proposed in \KAZMIG\ in the form of the matrix quantum
mechanics (MQM) of \BIPZ.  The string theory has been constructed as
the collective field theory (the Das-Jevicki-Sakita theory), in which
the only excitation, the massless "tachyon", was related to the
eigenvalue density of the matrix field.

In the framework of MQM it is difficult to formulate directly a string
theory in a nontrivial background metric since the operators
perturbing the metric do not have a simple realization.  However, we
can perturb the theory by other operators, like tachyon or winding
operators. We hope that such a perturbation can also produce a curved
metric but in an indirect way.  An example of such a duality was given
in \FZZ\ where the 2D black hole background is
induced by a winding mode perturbation.

 The 2D string theory in the simplest, translational invariant
background (the linear dilaton background) is described by the singlet
sector of MQM with inversed quadratic potential.  In the singlet sector
of the Hilbert space, the eigenvalues have Fermi statistics, and the
problem of calculating the $S$-matrix of the string theory tachyon
becomes a rather standard problem in a quantum theory of a
one-dimensional nonrelativistic free fermionic field.  The tree-level
$S$-matrix can be extracted by considering the propagation of "pulses"
along the Fermi surface and their reflection off the "Liouville wall"
\refs{\POLCHINSKI,\JEVICKI}.

The very existence of a formulation in terms of free fermions
indicates that the 2D string theory should be integrable.  For
example, the fermionic picture allows to calculate the exact
$S$-matrix elements. Each $S$-matrix element can be associated with a
single fermionic loop with a number of external lines \MPR.  One can
then expect that the theory is also solvable in a nontrivial,
time-dependent background generated by a finite tachyonic source.
Dijkgraaf, Moore and Plesser \DMP\ demonstrated that this is indeed
the case when the allowed momenta form a lattice as in the case of the
compactified Euclidean theory.  In \DMP\ it has been shown that the
string theory compactified at any radius $R$ possesses the integrable
structure of the Toda lattice hierarchy.  However, this important
observation had not been, until recently, really exploited.  The Toda
structure is too general, and it becomes really of use only if
accompanied by an initial condition or a constraint (a "string
equation"), which eliminates all the solutions but one. Thus the first
results for a nontrivial background in case of a general radius were
obtained as a perturbative expansion in the tachyon source \MOORE.

Recently, the Toda integrable structure of the compactified Euclidean
2D string theory was rediscovered \refs{\HKK,\KKK} and used to find the
explicit solution of the theory \refs{\KKK, \AK}. Finally, the string
equation at an arbitrary compactification radius was found in \IK.
These papers studied the T-dual formulation of the string theory,
where instead of the discrete spectrum of tachyon excitations the
winding modes of the string around the compactification circle
(Berezinski-Kosterlitz-Thouless vortices) were used.  These modes
appear in the non-singlet sectors of MQM \GRKL\ and can be generated
by imposing twisted periodic boundary conditions \BULKA.

In this paper we return to the study of the Toda integrable structure
of tachyon excitations of 2D string theory originated in \DMP.
It describes special perturbations of the ground state within the singlet
sector of the MQM.  We will construct the Lax operators as operators
in the Hilbert space of the singlet sector of MQM.  We will be able to
find an interpretation of the Toda spectral parameters in terms of the
coordinates of the two-dimensional target phase space and interpret
the solutions of the Toda hierarchy in terms of the shape of the Fermi
sea. In particular, we find the explicit shape of the Fermi sea for
the Sine-Liouville string theory.

We can give two different interpretations of our problem.  The first
one is that of a 2D string theory in Minkowski space in a
non-stationary background.  The simplest, time-independent ground
state of the theory is characterized by a condensation of the constant
tachyon mode, which is controlled by the cosmological constant $\mu$.
Here we will study more general, time-dependent backgrounds
characterized by a set of coupling constants $t_{\pm n}$ associated
with non-trivial tachyon modes with purely imaginary energies $E_n = i
n/R$, where $R$ is a real number.  Since the incoming/outgoing
tachyons with imaginary energies have wave functions exponentially
decreasing/increasing with time, such a ground state will contain only
incoming tachyons in the far past and only outgoing tachyons in the
far future.  In other words, the right and left vacua are replaced by
coherent states depending correspondingly on the couplings $t_n$ and
$t_{-n}$ $(n=1,2,...)$, which will be identified with the ``times'' of
the Toda hierarchy. The coherent states of bosons modify the
asymptotics of the profile of the Fermi sea at far past and future,
without changing the number of fermions.  The flow of the fermionic
liquid is no more stationary, but its time dependence happens to be
quite trivial, and the profiles of the Fermi sea at different moments
of time are related by Lorentz boosts.

The second interpretation of our analysis is that we consider
perturbations of an Euclidean 2D string theory in which the Euclidean
time is compactified at some length $\beta = 2\pi R$. Together with
the cosmological term, we allow all possible vertex operators with
momenta $n/R$, $n\in \Z$.  The simplest case of such a perturbation is
the Sine-Liouville theory.  This case will be considered in detail and
the shape of the Fermi sea produced by the Sine-Liouville perturbation
will be found explicitly. The solution we have found exhibits
interesting thermodynamical properties, which may be relevant to the
thermodynamics of the string theory on the dilatonic 2D black hole
background.

Let us mention also another possible application of our analysis: the
fermionic system that appears in MQM is similar to the problem of
two-dimensional fermions in a strong transverse magnetic field
\FERTIGone.  The electrons filling the first Landau level form
stationary droplets of Fermi liquid, similar to that of the
eigenvalues in the phase space of MQM.  The form of such droplets is
also described by the Toda hierarchy \WIEGAGAM. Our problem might be
related to the situation in which two such droplets are about to
touch, which can happen at a saddle point of the effective potential
\FERTIGtwo.

The paper is organized as follows. In section 2 we will remind the CFT
formulation of the 2D string theory. In section 3 the MQM in the
inverted harmonic potential in the ``light-cone'' phase space
variables is formulated and the free fermion structure of its singlet
sector is revealed. In section 4 the one particle wave functions are
studied. In section 5, after the description of the fermionic ground
state, tachyonic perturbations are introduced and the equations
defining the corresponding time-dependent profile of the Fermi sea are
derived. In section 6 the Toda integrable structure of the
perturbations restricted to a lattice of Euclidean momenta is derived
directly from the Schroedinger picture for the free fermions. In
section 7 we recover the solution of the Sine-Liouville model and
describe explicitly its Fermi sea profile.  In section 8 we reproduce
the free energy of the perturbed system from the profile of the Fermi
sea. The section 9 is devoted to conclusions and in section 10 we
discuss some problems and future perspectives. In particular, we
propose a 3-matrix model describing the 2D string theory perturbed by
both tachyon and winding modes. In the appendix the one particle wave
functions of the type II model, defined on both sides of the quadratic
potential,  in the ``light cone'' formalism are presented.

\newsec{Tachyon and winding modes in 2D Euclidean string theory}

The 2D string theory is defined by Polyakov Euclidean string action
\eqn\PSTR{
S(x,g)={1\over 4\pi}\int d^2\sigma\sqrt{{\rm det} g}
[ g^{ab}\p_a x\p_b x +\mu],  }
where the bosonic field $x(\sigma)$ describes the embedding of the
string into the Euclidean time dimension and $ g^{ab}$ is a world
sheet metric.  In the conformal gauge $g^{ab}=e^{\phi(\s)}\hat
g^{ab}$, where $\hat g^{ab}$ is a background metric, the dilaton field
$\phi$ becomes dynamical due to the conformal anomaly and the
world-sheet CFT action takes the familiar Liouville form
\eqn\confstr{
S_0={1\over4\pi}\int _{\rm world\  sheet} d^2 \sigma\, [(\partial x)^2
+(\partial\phi)^2 -4\hat \CR\phi + \mu e^{-2\phi}+{\rm ghosts}].}
This action describes the unperturbed linear dilaton string
background corresponding to the flat 2D target space $(x,\phi)$.

In the target space this theory possesses only one propagating degree
of freedom which corresponds to tachyon field.  If the Euclidean time
is compactified to a circle of radius $R$, $x(\sigma)\equiv
x(\sigma)+2\pi R$, the spectrum of admissible momenta is discrete:
$p_n=n/R,\ \ n\in \Zb$.  In this case there is also a discrete
spectrum of vortex operators, describing the winding modes
(Kosterlitz-Thouless vortices) on the world sheet. A vortex of charge
$q_m=mR$ corresponds to a discontinuity $2\pi m R$ of the time
variable around a point on the world sheet.\foot{At rational values of
$R$ there exist additional physical operators (similar to discrete
states at the self-dual radius $R=1$) containing the derivatives of
fields $x(\s)$ and $\phi(\s)$, which we will ignore in this paper.}
The explicit expressions of the vertex operators $V_p$ and the vortex
operators $\tilde V_q$ in terms of the position field $x=x_R+x_L$ and
its dual $\tilde x=x_R-x_L$ are
\eqn\vert{\eqalign{
V_p &= {\Gamma(|p|)\over \Gamma(-|p|)}\int d^2
 \sigma\, e^{-i p x }e^{( |p|-2)\phi},\cr
\tilde  V_q& = {\Gamma(|q|)\over \Gamma(-|q|)}
\int d^2 \sigma\,  e^{-i q \tilde x }e^{(|q| -2)\phi}. } }

 Any background of the 
compactified 2D string theory can be obtained
(at least in the case of irrational $R$) by a perturbation of the
action \confstr\ with both vertex and vortex operators
\eqn\PERTS{ S=S_0+\sum_{n\ne 0}( t_n V_n +\tilde t_n\tilde V_n).  } 
Such a theory possesses the T-duality symmetry of the non-perturbed
theory $x\leftrightarrow \tilde x$, $R\to1/R$, $\mu\to \mu/R$
\KLEBANOV\ if one also exchanges the couplings as
$t_n\leftrightarrow\tilde t_n$.  Another general feature of the theory
\PERTS\ is the existence of the physical scaling of various couplings,
including the string coupling $g_s$, with respect to the cosmological
coupling $\mu$.  It can be found from the zero mode shift of the
dilaton $\phi\to\phi+\phi_0$. In this way we obtain
\eqn\FSCAL{ 
t_n\sim \mu^{1-\hf|n|/R},\quad
\tilde t_n\sim\mu^{1-\hf R|n|},\quad   
 g_s\sim\mu^{-1}.  }

The scaling \FSCAL\ allows to conclude that at the compactification
radius lower than the Berezinski-Kosterlitz-Thouless radius
$R_{KT}=1/2$ all vertex operators are irrelevant.  In the interval
$1/2<R<1$ the only relevant momenta are $p= \pm 1/R$. The theory
perturbed by such operators looks as the Sine-Gordon theory coupled to
2d gravity (``Sine-Liouville'' theory):
\eqn\SGSL{ 
S_{SG}={1\over4\pi}\int d^2 \z\, [(\partial x)^2
+(\partial\phi)^2 -4\hat \CR\phi  + \mu e^{-2\phi} +
\l e^{({1\over R} -2)\phi} \cos(  {x}/R )]. }
It was conjectured by \FZZ\ that at $R=2/3$ and
$\mu=0$ the ``Sine-Liouville'' is dual to the $\left[{SL(2,C)\over
SU(2)\times U(1)}\right]_k$ coset model with central charge 
$c={3k\over k-2}-1=26$, which describes the 2D string theory in
the black hole (``cigar'') background.

If we go to the Minkowski space, the perturbation \PERTS\ is made by
tachyons with purely imaginary momenta.  In this case there are more
general perturbations, generated by tachyons with any real energies.
In the next section we will introduce the matrix formulation of the
string theory in Minkowski space using light-cone coordinates.

\newsec{Matrix Quantum Mechanics in light cone formulation}

The 2D string theory in Minkowski time appears as the collective field
theory for the large-$N$ limit of MQM in the inverted oscillator
potential
\refs{\KAZMIG,\BRKA,\PARISI,\GRMI,\GIZI}.  The matrix Hamiltonian is
\eqn\HamO{
H= \hf \Tr (P^2 - M^2), }
where $P=-i{\p/ \p M}$ and $M_{ij}$ is an $N\times N$ hermitian
matrix variable. The cosmological constant $\mu$ in 
\confstr\ is introduced as a ``chemical potential'' 
coupled to the size of the matrix $N$,
which should be considered as a dynamical variable.  
In this formulation the time coordinate of the string target space
coincides with the MQM time (or its Euclidean analogue $\xx=- it$ in
the compactified version of the MQM) and that of the Liouville field
is related to the spectral variable of the random matrix.

The tachyon modes of the string theory are represented by the asymptotic states of the collective theory. The scattering operators with real energy $E$ and
describing left- and right-moving waves, respectively, are given by
  \JEVICKI\eqn\Tachopp{ \eqalign{ T_+^E     =e^{-iEt} \Tr (M+ P)^{iE},  
\qquad T_-^E = e^{-iEt} \Tr (M-P)^{-iE}}.}
These operators can be used to construct the in- and out-states of scattering theory.  Namely, for an in-state, one needs a left-moving wave while the out-state is necessarily given by a right-moving one.  
 The vertex operators \vert\ are tachyons with purely imaginary energies
and are therefore
represented by the following chiral operators in the MQM
\eqn\Vopp{ {V _p\to  \cases{ e^{-pt}\Tr  (M+P)^{|p|},  &  $p> 0$, \cr
 e^{-pt}\Tr  (M-P)^{|p| },& $p< 0$. }}}
Since we are interested in perturbations with the chiral
operators \Vopp, it is natural to perform a canonical transformation to
light-cone variables
\eqn\YpYm{\Xpm = {M\pm P\over \sqrt{2}}}
and write the matrix Hamiltonian as
\eqn\hammat{
H_0 = -  \hf \Tr ( \Xp \Xm+  \Xm\Xp),}
where the matrix operators $\Xpm$ obey the canonical commutation relation 
\eqn\ccrYY{ [(\Xp)^i_j,(\Xm)^k_l]=-i  \ \delta^{i}_{l}\delta^k_j.}
Define the right and left Hilbert spaces as the spaces of functions 
of $\Xp$ and $\Xm$ only, with the scalar product
\eqn\SCPR{ \langle \Phipm|\Phipm'\rangle 
=\int d^{N^2}\Xpm \, \overline{ \Phipm(\Xpm)}\Phipm'(\Xpm).
}
The operator of coordinate in the right Hilbert space is 
the momentum operator in the left one and the wave functions 
in the $\Xp$ and $\Xm$ representations are related by a Fourier transform.
The second-order Schr\"odinger equation associated with the Hamiltonian 
\HamO\
%
%
becomes a first-order one when written in the $\pm$-representations  
\eqn\SCHRPM{
 \p_t \Phipm(\Xpm,t)= \mp \Tr \left(\Xpm {\p\over\p \Xpm}
 +{N \over 2} \right)\Phipm(\Xpm,t), }
whose general solution is
\eqn\GENSOL{ \Phipm (\Xpm , t)  =   
 e^{\mp \hf N^2 t}  \Phipm^{(0)}(e^{\mp t}\Xpm ).}

The Hilbert space decomposes into a direct
sum of eigenspaces labeled by the irreducible representations of $SU(N)$,
which are invariant under the action of the Hamiltonian \hammat.
The functions $ \vec \Phi^{(r)}_\pm=\{\Phi^{(r, J)}_\pm\}_{J=1}^{{\dim} (r)}$ 
belonging to given irreducible representation $r$ transform as
\eqn\IRREP{ \Phi_\pm^{(r,I)}(\O^\dagger\Xpm\O)=
\sum_{J}\gr(\O)\Phi^{(r,J)}_\pm(\Xpm), }
where $\gr$ is the group matrix element in representation $r$ and
$I,J$ are the representation indices.  The wave functions transforming
according to given irreducible representation depend only on the $N$
eigenvalues $\zipm{1},\dots,\zipm{N}$ and the Hamiltonian \hammat\
reduces to its radial part
\eqn\ANGH{ H_0 = \mp i \sum_k (\zkpm\p_{\zkpm}+ {N\over 2}). }
 A potential advantage of the light-cone approach is that the
Hamiltonian \ANGH\ does not contain any angular part, which is not the
case for the standard one \HamO, whose angular term induces a
Calogero-like interaction.

In the scalar product \SCPR, the angular part can also be integrated
out, leaving only the trace over the representation indices:
\eqn\SCPRO{ \langle \vec \Phpm{(r)}|\vec \PPhpm{(r)}\rangle=\sum_J
\int \prod_k d\zkpm  \Delta^2(\xpm)
\overline{\Phi^{(r,J)}_\pm(\xpm)}\Phi'^{(r,J)}_\pm(\xpm),  }
where $\Delta(\xpm)$ is the Vandermonde determinant.
If we define
\eqn\FWF{\vec  \Pspm^{(r)}(\xpm) =\Delta(\xpm)\vec\Phpm{(r)}(\xpm)   ,  }
these determinants disappear from the scalar product.
The Hamiltonian for the new functions $\vec \Pspm^{(r)}(\xpm)$ 
takes the same form as \ANGH\ but with a different constant term:
\eqn\ANGHD{  H_0 = \mp i \sum_k (\zkpm\p_{\zkpm}+ {1/2}). }

 In the singlet representation, the wave function $\Psi_{\pm}
(\xpm)\equiv \Psi_\pm^{({\rm singlet} )}(\xpm)$ is a completely
antisymmetric scalar function. The scalar product in the singlet
representation is given by
\eqn\SCPRO{ \langle \Psi_\pm |\Psi_\pm '\rangle=
\int \prod_k d\zkpm 
\overline{\Psi_\pm (\xpm)}\Psi_\pm '(\xpm)  . } 
Thus the singlet sector describes a system of $N$ free fermions. 
The singlet eigenfunctions of the Hamiltonian \ANGHD\ are
represented by Slater determinants of one-particle eigenfunctions
discussed in the next section.

It is known \refs{\GRKL,\BULKA,\KKK} that unlike singlet, which is
free of vortices, the adjoint representation contains string states
with a vortex-antivortex pair, and higher representations contain
higher number of such pairs.  In what follows we will concentrate on
the fermionic system describing the singlet sector of the matrix
model. We will start from the properties of the ground state of the
model, representing the unperturbed 2D string background and then go
over to the perturbed fermionic states describing the (time-dependent)
backgrounds characterized by nonzero expectation values of some vertex
operators.

\newsec{Eigenfunctions and fermionic scattering}

To study the system of non-interacting fermions we have to start with
one-particle eigenfunctions. The one-particle Hamiltonian in the
light-cone variables of the previous section can be written as
\eqn\oneph{H_0
= -\hf (\hat \xp\hat \xm +\hat \xm\hat \xp),}
where $\xpm$ turn out to be canonically conjugated variables 
\eqn\ygrekpm{
[ \hat \xp,\hat \xm]=-i.}
  We can work either in $\xp$ or in $\xm$
representation, where the theory is defined in terms of fermionic
fields $\psipm(\xpm)$ respectively. 
General solutions of the Schr\"odinger equation with the Hamiltonian \oneph\
written in these representations take the form
\eqn\GENSOL{ \psipm (\xpm , t)  =   
\xpm^{-1/2}  \ f_{_{\pm}} (e^{\mp t}\xpm ).}

There are two versions of the theory, referred in \MPR\ as theories of
type I and II. In the theory of type I the eigenvalues $\l_k$ of the
original matrix field $M$ are restricted to the positive half-line.
The theory of type II is defined on the whole real axis $\l$. Such a
string theory splits, at the perturbative level, into two disconnected
string theories of type I.  Here we will consider the fermion
eigenfunctions in the theory of type I. The theory of type II will be
considered in Appendix A.
 
In the light cone formalism it is natural to replace the restriction
$\l>0$ with $\xpm>0$, which again does not affect the perturbative
behavior.  In this case the solutions with a given energy are
$\psepm(\xpm,t)= e^{- iEt} \psepm(\xpm)$ with
\eqn\wavef{   \psepm(\xpm)=
{1\over\sqrt{2\pi}}
\xpm^{\pm iE-\hf}.
}
 The functions \wavef\ with $E$ real form two complete systems of
$\delta$-function normalized orthonormal states
\eqn\normpm{\langle \psepm|\peepm\rangle\equiv
 \int_{0} ^{\infty} d\xpm\,
\overline{\psepm(\xpm)}\peepm(\xpm) =
\delta(E -E'),
} 
\eqn\complpm{\int_{-\infty}^{\infty}dE\,  \ 
\overline{\psepm(\xpm)}\psepm(\xpm')= \delta( \xpm-\xpm').
}

As any two representations of a quantum
mechanical system, the $\xp$ and $\xm$ representations 
are related by a unitary operator
$\hat S$, which in our case is nothing but the Fourier transformation 
on the half-line.  The latter is defined 
by the integral
\eqn\Fourtr{[\hat S \psip](\xm)=  \int_{0}^{\infty}
d\xp \, \KK(\xm,\xp) \psip(\xp),}
where there are two choices for the kernel:
\eqn\ker{\KK(\xm,\xp)=\sqrt{2\over \pi}\cos(\xm\xp)\ \  {\rm or} \ \
\KK(\xm,\xp)=i\sqrt{2\over \pi}\sin(\xm\xp). } 
The sine and the cosine kernels describe two possible theories, 
which differ only on the non-perturbative level.\foot{The fact that there are two choices for the kernel can
be explained as follows. In order to define the theory of type I
for the original second-order Hamiltonian
\HamO, we should  fix the boundary condition at $\l=0$, and there are 
two linearly independent boundary conditions.}  Let us choose the
cosine kernel in \ker\ and evaluate the action of the $\hat
S$-operator on the eigenfunctions.  This integral is essentially the
defining integral for the $\Gamma$-function:
\eqn\Smtrxx{[\hat S^{\pm 1} \psepm ](\xmp) =  {1\over \pi}
\int_{0}^{\infty} d\xpm\,  \cos( \xp\xm)  
  \xpm^{\pm iE-\hf} 
= \CR(\pm E) \psemp (\xmp), }
 where
\eqn\rfactor{
\CR(E) = \sqrt{2\over \pi}
\cosh\left({\pi\over 2} (i/2-E)\right)  \Gamma( iE + 1/2). 
} 
(The sine kernel would give the same result, but with cosh replaced by sinh).
The factor $\CR(E)$ is a pure phase  
\eqn\unitrtyr{ \overline{\CR(E) } \CR(E) = \CR(-E) \CR(E)=1,}
which proves the unitarity of the operator $\hat S$.  The operator
$\hat S$ relates the incoming and the outgoing waves and therefore can
be interpreted as the fermionic scattering matrix. The factor $\CR(E)$
is identical to the the reflection coefficient (the "bounce factor" of
\MPR), characterizing the scattering off the upside-down oscillator
potential.  In the standard scattering picture, the reflection
coefficient is extracted by comparing the incoming and outgoing waves
in the large-$\l$ asymptotics of the exact eigenfunction of the
inversed oscillator Hamiltonian $H= -\hf(\p_\l^2 +\l^2)$.

It follows from the above discussion that the scattering amplitude
between two arbitrary in and out states is given by the integral with
the Fourier kernel \ker
\eqn\scpr{\eqalign{
& \langle\psim|\hat S\psip\rangle  =\langle\hat S^{-1}
\psim|\psip\rangle =
\langle\psim|K|\psip\rangle \cr
& \langle\psim|K|\psip\rangle 
\equiv
\int_{0}^\infty d\xp d\xm\,
\overline{\psim(\xm)} \   \KK(\xm,\xp)      \psip(\xp).
}} 
The  integral \scpr\ can be interpreted as a scalar product between the in and out states. 
By \normpm\ and \Smtrxx\ one finds that the in and out eigenfunctions satisfy the orthogonality relation
\eqn\scprEE{
\langle\psem|K|\peep\rangle= \CR(E)\delta(E-E').}

\newsec{String theory backgrounds as profiles of the Fermi sea}

The ground state of the  MQM  is obtained by filling all energy
levels up to some fixed Fermi energy which we choose to be $E_F=-\mu$.
Quasiclassically every energy level corresponds to a certain
trajectory in the phase space of $\xp,\xm$ variables.  
The Fermi sea can be viewed as a stack of all classical trajectories
with $E<E_F$ and the ground state is completely
characterized by the curve representing the trajectory of the 
fermion with highest energy $E_F$. For the Hamiltonian \oneph\
all trajectories are hyperboles $\xp\xm = -E$ 
and the profile of the  Fermi sea is given by 
\eqn\TRAJ{ \xp\xm=\mu . }

In the quasiclassical limit the phase-space density of fermions is
either 0 or 1. Then the low lying collective excitations are
represented by deformations of the Fermi surface, {\it i.e.}, is the
boundary of the region in the phase space filled by fermions.  At any
moment of time such deformation can be obtained by replacing the
constant $\mu$ on the right hand side of \TRAJ\ with a function of
$\xp$ and $\xm$
\eqn\eqpr{ \xp\xm = M(\xp, \xm).}
 In contrast to the ground state which is stationary, an excited state
given by a generic function $M$ leads to a time dependent
profile. However, this dependence is quite trivial: since the Fermi
surface consists of free fermions each moving according to its
classical trajectory $\xpm(t)=e^{\pm t}\zpm(0)$, where $\zpm(0)$ is
the initial data, we can always replace \eqpr\ by the equation for the
initial shape.  So the evolution of a shape in time is simply its
homogeneous extension with the factor $e^{t}$ along the $\xp$ axis and
a homogeneous contraction with the same factor along the $\xm$ axis.

Below we will find equations that determine the shape of the Fermi sea
for a generic perturbation with tachyon operators.  Our analysis is in
the spirit of the Polchinski's derivation of the tree-level $S$-matrix
\POLCHINSKI, but we will consider finite, and not only infinitesimal
perturbation.

  In terms of the Fermi liquid, the incoming and the outgoing states
are defined by the asymptotics of the profile of the Fermi sea at
$\xp\gg\xm$ and $\xm\gg\xp$, correspondingly. If we want to consider
such a perturbation as a new fermion ground state, we should change
the one-fermion wave functions. The new wave functions are related to
the old ones by a phase factor
\eqn\asswave{
\Pepm(\xpm)=e^{\mp i \vp_\pm(\xpm;E)}
\psepm(\xpm), }
whose asymptotics at large $\xpm$ characterizes the incoming/outgoing
tachyon state. We split the phase into three terms
\eqn\pot{
\vp_\pm  (\xpm;E) 
= V_\pm(\xpm) +\hf \phi(E) + v_\pm(\xpm;E),}
 where the potentials $V_\pm$ are fixed smooth functions vanishing at
$\xpm=0$, while the term $v_\pm$ vanishing at infinity and the
constant $\phi$ are to be determined.  Thus, the potentials $V_\pm$
fix unambiguously the perturbation.  The time evolution of these
states with the original Hamiltonian \oneph\ is determined by the
eq. \GENSOL. To see that the state \asswave\ contains incomming and
outgoing tachyons, it is enough to note that it arises as a coherent
state of vertex operators \Vopp\ acting on the unperturbed wave
function \wavef.

Since the functions $\Pep$ and $\Pem$ should describe the same
one-fermion state, the Fourier transform \Fourtr\ should be diagonal
in their basis. We fix the zero mode $\phi$ so that
\eqn\PepSm{
\hat S \ \Pep= \Pem.}
This condition can also be expressed as the orthonormality of in and
out eigenfunctions
\asswave\  %
\eqn\orth{
\langle  \Pim{E_{{-}}}|K|  \Pip{E_{{+}}}\rangle =\delta(E_+-E_-),}
with respect to the scalar product \scpr.  This requirement fixes the
exact form of the wave functions.  Let us look at this problem in the
quasiclassical limit $E_\pm \to \infty$.  The scalar product in \orth\
is written as
\eqn\scort{
\langle \Pim{E_{{-}}}|\Pip{E_{{+}}}\rangle={e^{-i\phi}
\over \ \pi \sqrt{2\pi}}
\int\limits_{0}^\infty {d\xp d\xm\over \sqrt{\xp\xm} }\cos(\xp\xm) 
e^{- i \vp_+(\zp)-i\vp_-(\zm) } 
\xm^{iE_{{-}}}\xp^{iE_{{+}}} .
} At the quasiclassical level it can be evaluated by the saddle point
approximation. One obtains two equations for the saddle point\foot{We
assume that only one exponent of $\cos(\zp\zm)$ gives a
contribution. The result will be shortly justified from another point
of view.}
\eqn\sdlpt{ 
\zp\zm =-E_\pm+\zpm \p \vp_\pm(\xpm).}
Generically, the two equations \sdlpt\ define two different curves in
the $\zp$-$\zm$ plane, and their compatibility can render at most a
discrete set of saddle points.  However, if the two solutions define
the same curve, we obtain a whole line of saddle points, which implies
the existence of a zero mode in the double integral \scort.  The
contribution of the zero mode is infinite, amounting to the
$\delta$-functional orthogonality relation. Thus, the orthonormality
condition is reduced to the compatibility of two equations \sdlpt\ at
$E_+=E_-$.

For example, in the absence of perturbations ($\vp_\pm=0$), the saddle
point equations $\xp\xm=-E_\pm$ are inconsistent, unless
$E_+=E_-$. For equal energies they coincide
leading to a zero mode in the double integral \scprEE. The resulting
saddle point equation gives  the classical hyperbolic trajectory in the phase
space of an individual fermion.  If $E_+\ne E_-$, the integrand of
\scort\ is a rapidly oscillating function and the integral is zero.

The equations \sdlpt\ and the requirement of their compatibility can
be obtained from another point of view. The perturbed wave function
\asswave\ can be interpreted as an eigenstate of a new, perturbed
Hamiltonian. Indeed, the functions \asswave\ are not eigenstates of
the original Hamiltonian \oneph\ in $\pm$-representations $H_0^\pm=
\mp i(
\xpm \p_{\xpm} +1/2) $, but for each given $E$
they are evidently eigenstates of the operators  
\eqn\Hpertu{H_\pm^{_E}=  H_0^\pm+ \xpm \p \vp_{\pm}(\xpm;E),}
where $\vp_{\pm}$ contain so far unknown functions $v_{\pm}(\zpm;E)$.
However, the operators $H_\pm^{_E}$ depend on the energy $E$ through
these functions. Therefore, they can not be considered as
Hamiltonians.  But one can define the Hamiltonians as solutions of the
equations
\eqn\Hper{H_\pm=  H_0^\pm+ \xpm \p \vp_{\pm}(\xpm;H_\pm).}
Then all functions \asswave\ are their eigenstates. 

The orthonormality condition \orth\ can be equivalently rewritten as the
condition that the Hamiltonians $H_\pm$ define the action of the 
same self-adjoint operator $H$ in the $\pm$-representations.
 In the quasiclassical limit, this is equivalent to
the coincidence of the  phase space trajectories  associated with 
$H_+$ and $H_-$:
\eqn\eqper{
H_+(\zp,\zm)=E \Leftrightarrow H_-(\zp,\zm)=E.  }  
This condition is equivalent to the compatibility of two equations
\sdlpt.

Note that with respect to the time $\tau$ defined by the new
Hamiltonian, the time dependence of the states characterized by the
wave functions \asswave\ is given by $e^{-iE\tau}$.  This corresponds
to a stationary flow of the Fermi liquid. The profile of the Fermi sea
coincides with the classical trajectory of the fermion with the
highest energy $E_F=-\mu$. Its equation can be written in two forms
similar to \eqpr\ which should be consistent
\eqn\TOD{ 
\zp\zm =M_\pm(\zpm)\equiv \mu +\zpm \p \vp_\pm(\xpm;-\mu).}
These equations are the non-compact version of the equations arising
in the conformal map problem which is a semiclassical description of
compact Fermi droplet \WIEGAGAM. In that case the potential must be an
entire function which is not required in our case. The functional
equation \TOD\ contains all the information of the tachyon
interactions in the tree level string theory. To proceed further with
our analysis, we should concretize the form of the perturbing
potentials $V_\pm$. We will show in the next section that the
perturbations produced by vertex operators are integrable.

\newsec{Integrable flows associated with  vertex operators}

\subsec{Lax formalism, Toda Lattice structure and string equations}

Now we restrict ourselves to the time dependent coherent states made of
tachyons with discrete Euclidean momenta $p_n=n/R$. This spectrum of
momenta arises when the system compactified on a circle of length
$2\pi R$, or heated to the temperature $T=1/(2\pi R)$. Such perturbations
are described by the potentials of the following form
\eqn\Vbig{V_\pm(\xpm)
= R\sum\limits_{k\ge 1} t_{\pm k} \xpm^{k/R} . }
In this section we will show that such a deformation is exactly solvable,
being generated by a system of commuting flows
$H_n$ associated with the coupling constants $t_{\pm n}$.
The associated integrable structure is that of  a constrained Toda Lattice
hierarchy. The method
is very similar to the standard Lax formalism of Toda theory, 
but we will not assume that the reader is familiar with
this subject. It is based on rewriting all operators in
the energy representation, in which the Fourier transformation $\hat S$ is
diagonal as we required in the previous section. The energy $E$ will 
play the role of the coordinate along the lattice formed by the allowed 
energies  $E_n = ip_n$ of the   tachyons. 
       
Let us start with the operators $\hat \xpm = \xpm$ 
and $ \hat \p_\mp= {\p\over\p \xmp}$, which 
are related as
\eqn\FoirO{-i \hat \p_-=   \hat   S \hat \xp \hat S^{-1},
\qquad  i \hat \p_+=   \hat   S^{-1} \hat \xm \hat S.}
The Heisenberg commutation relation $[\hat \p_\pm , \hat \xpm]=1$ 
leads to the operator identity
\eqn\streQ{\eqalign{
\hat \xm \hat   S  \hat \xp \hat S^{-1} - 
 \hat S \hat \xp \hat   S^{-1}  \hat \xm= i.
}}
Further, the Hamiltonian \oneph\ in the $\xpm$-representation 
$H^\pm_0= \mp i(\xpm \p_{\xpm} + 1/2 )$  is expressed 
in terms of $\hat \xpm$ and $\hat S$ as
\eqn\HamOp{ 
H_0^- =- \hf\left(\hat \xm \hat S  \hat \xp\hat S^{-1} 
+\hat S\hat \xp \hat   S^{-1} \hat \xm\right)=
\hat S H_0^+ \hat S^{-1} .
}

It follows from the identities
\eqn\psiHzz{ \eqalign{
H_0^\pm \psepm(\xpm) &= E \psepm(\xpm) , \cr
\xpm \psepm (\xpm) & = \psi_\pm^{_{E\mp i} }(\xpm), \cr
 \hat S ^{\pm 1}\psepm(\xpm) &= \CR^{\pm 1}(E)\psemp(\xmp) }}
 that the
operators $H_0^\pm$, $\hat S$ and $\hat \xpm$ are represented in the
basis of the non-perturbed wave functions \wavef\ by
\eqn\bareREP{ 
H_0^\pm = \hat E,  \qquad 
\hat\xpm  = \oo^{\pm 1},  \qquad  
\hat S = \CR(\hat E)\iota \equiv \hat \CR \iota,
}
where $\iota\psepm=\psemp$ and $\oo$ is the shift operator 
\eqn\oomega{\oo=
e^{-i\p_E}.}  In order to have a closed operator algebra, we should
also define the action of the operators $\hat S^{\pm 1} $ on the
functions $\psi_\pm^{_{E\mp n i} }(\xpm)= \xpm^n \psi_\pm^{_{E}
}(\xpm) $.  To satisfy the identities \FoirO , we should define $\hat
S^{\pm 1}$ by the cosine kernel for $n$ even and by the sine kernel
for $n$ odd.  Therefore $ \hat S \psi_\pm^{_{E\mp i} }(\xpm)= \hat
\CR'
\psi_\pm^{_{E\mp  i} }(\xpm)$
where $\hat \CR' = \CR'(\hat E)$ is given by 
\rfactor\ with cosh replaced by sinh.

The algebraic relations \streQ\ and \HamOp\ are transformed into
algebraic relations among the operators $\hat E$, $\hat \o$ and
$\hat\CR$ in the $E$-space\foot{Note that we should inverse the order
of the operators.}
\eqn\StreQ{\eqalign{
\hat\CR^{-1} \oo \hat\CR' \oo ^{-1}   &= -\hat E+i/2,\cr
\oo^{-1} \hat\CR'^{-1}\oo\hat\CR&=-\hat E -i/2. }}
This is equivalent, according to the remark above, to the functional
constraint
\eqn\funcE{\CR(E)=- (i/2 +E) \ \CR'(E+i),}
which is evidently satisfied.  To simplify the further discussion, in
the rest of this section we will identify the functions $\CR(E)$ and
$\CR'(E) $, thus neglecting the non-perturbative terms $\sim e^{\pi
E}$.  Note however that all statements made below can be proved
without this identification.

Our aim is to extend the $E$-space representation for the basis of
wave functions \wavef\ perturbed by a phase factor $e^{\pm i\vp_\pm}$
as in eqs. \asswave-\pot, with $V_\pm$ given by \Vbig.  We assume that
the phase $\vp_\pm$ can be expanded, for sufficiently large $\xpm$ as
a Laurent series
\eqn\VPET{ \vp_\pm (\xpm) = R\sum\limits_{k\ge 1} t_{\pm k} \xpm^{k/R}
 +\hf \phi  -R\sum\limits_{k\ge 1} {1\over k} v_{\pm k}
\zpm^{-k/R}.}
This assumption will be justified by the subsequent analysis.
With the special form \VPET\ of the perturbing phase, 
there exist unitary ``dressing operators" $\CW_\pm$
acting in  the $E$-space, which transform the ``bare" wave functions
$\psepm$ into the ``dressed" ones
\eqn\Drss{ \Pepm\equiv e^{\mp i \vp_\pm(\xpm) }\psepm=\CW_\pm
\psepm.}
It is evident from \VPET\ and \bareREP\ that the dressing operators
can be expressed as series in the shift operator $\oo$ with
$E$-dependent coefficients
\eqn\DessOp{
\CW_\pm =e^{\mp i\phi/2}\ 
\left( 1+ \sum_{k\ge 1} w_{\pm k} \hat \o^{\mp k/R}\right)
e^{\mp i R\sum_{k\ge 1} t_{\pm k} \hat \o ^{\pm k/R}}.}

In the basis of the perturbed wave functions \Drss, the operators of
canonical coordinates $\hat \xpm$ are represented by two Lax operators
\eqn\LpLmG{ L_\pm = \CW_\pm
\ \oo^{\pm 1}\ \CW^{-1}_\pm = e^{\mp i\phi/2} \oo^{\pm 1}e^{\pm i \phi/2} 
\left(1 +\sum_{k\ge 1} a_{\pm k}   \oo^{\mp k/R}\right)
}
 and the Hamiltonians $H_0^\pm$ are represented (up to a change of
sign) by the so-called Orlov-Shulman operators
\eqn\OrlSh{
M_\pm = -  \CW_\pm \hat E \CW^{-1}_\pm=
   \sum _{k\ge 1}  k t_{\pm k}   L_\pm^{ k/R}- \hat E +
\sum_{k\ge 1} v_{\pm k}  L_\pm^{-k/R}.
}
 In the above formulas all coefficients are functions of $E$ and
$\tv=\{t_{\pm k}\}_{k=1}^{\infty}$.  The Lax and Orlov-Shulman
operators act on the wave functions as
\eqn\LaxPepm{
L_\pm \Pepm(\zpm) = \zpm  \Pepm(\zpm),}
\eqn\OrlSPepm{
M_\pm \Pepm(\zpm) = \left(\sum _{k\ge 1}  k t_{\pm k}  \zpm^{ k/R}  - E+
\sum_{k\ge 1} v_{\pm k}  \zpm^{-k/R}\right)\Pepm(\zpm)
}
and satisfy the Lax equations
\eqn\LaxLM{
[L_\pm, M_\pm] = \pm  i L_\pm,}
which are the dressed version of
the relation  $[\oo^{\pm 1}, -\hat E]=\pm i \oo^{\pm 1}$.

The structure of constrained Toda hierarchy follows from the
requirement \PepSm, which means that the action of the $\hat S$
operator on the perturbed wave functions is totally compensated by the
dressing operators:
\eqn\TakTakT{    \CW_-=\CW_+ \hat \CR .}
The condition \TakTakT\ defines both the Toda structure and a
constraint which plays the role of an initial condition for the PDE of
the Toda hierarchy. 

The Toda structure implies that the tachyon operators generating the
perturbation are represented in the $E$-space by an infinite set of
commuting flows. To show this, we evaluate the variations of the Lax
operators with respect to the coupling constants $t_n$. From the
definition \LpLmG\ we have
\eqn\TodA{ 
\p_{t_n} L_\pm  = [ H_n, L_\pm], }
where the operators $H_n$ are related to the dressing operators as
\eqn\HWpm{
H_{n} = (\p_{t_ {n}}\CW_+) \CW_+^{-1} = (\p_{t_{n}}\CW_-) \CW_-^{-1}.}
The two representations of the flows $H_n$ are equivalent by virtue of
the relation \TakTakT. It is important that $\hat\CR$ does not depend
on $t_n$'s. A more explicit expression in terms of the Lax operators
is derived by the following standard argument.  Let us consider the
case $n>0$. From the explicit form of the dressing operators \DessOp\
it is clear that $H_n = W_+
\oo^{n/R} W_+^{-1} + $ negative powers of $\oo^{1/R}$. The variation
of $t_n$ will change only the coefficients of the expansions \LpLmG \
of the Lax operators, preserving their general form.  But it is clear
that if the expansion of $H_n$ contained negative powers of
$\oo^{1/R}$, its commutator with $L_-$ would create extra powers
$\oo^{-1 - k/R}$.  Therefore
\eqn\hashka{
 H_{ \pm n} =  (L_\pm^{n/R}   ) _{^{>}_{<} } +\hf (L_\pm^{n/R}   ) _{0} , 
\qquad n>0,
}
where the symbol $( \ \  )_{^{>}_{<}}$ means the positive
(negative) parts of the series in the shift operator
$ \oo $ and $( \ \ )_{ 0}$ means the constant part. 
 By a similar argument one shows that the Lax equations \TodA\
are equivalent to the zero-curvature conditions
\eqn\zerocur{ \p_{t_m} H_n -\p_{t_n}H_m - [H_m, H_n]=0.}
The equations \TodA\ and \hashka\ ensure that the perturbations
related with the couplings $t_n$ are described by the Toda integrable
structure.

The Toda structure leads to an infinite set of PDE's for the
coefficients $w_n$ of the dressing operators, the first of which is
the Toda equation for the zero mode of the dressing operators
\eqn\Todaeq{i{\p \over \p t_1}{\p \over \p t_{-1}} \phi
= e^{i\phi(E)-i\phi(E+i/R)} - e^{i\phi(E-i/R)-i\phi(E)} .  }
 The uniqueness of the solution is assured by appropriate boundary
conditions or additional constraints, known also as string equations.
In our case the string equation can be obtained by dressing the
equation \StreQ\ using the formula
\TakTakT,
which leads to the following relation between the Lax and
Orlov-Shulman operators
\eqn\STREQ{
  L_+ L_- = M_+ +i/2,\qquad
    L_- L_+=  M_- -i/2.
}
Similarly, the identity $\hat \CR \hat E = \hat E \hat \CR$ implies
\eqn\HamOp{ M_-=M_+.}
 
Thus, the perturbations of the one-fermion wave functions of the form
\Drss\ are described by a constrained Toda lattice hierarchy.  In the
T-dual formulation the same result was proved recently using the
standard definition of the Toda Lax operators \IK.  The standard Lax
and Orlov-Shulman operators are related to $L_\pm$ and $M_\pm$ by
\eqn\standLx{ L=L_+^{1/R}, \quad \bar L = L_-^{1/R}, \quad
M=M_+, \ \ \bar M=M_-.}
This operators satisfy $R$-dependent constraints \IK
\eqn\constrLx{
L^R \bar L^R = M +i/2, \ \ 
\bar L^R L^R = \bar M -i/2 , \ \ \ M=\bar M.
} 
(The string equations for integer values of $R$ have been discussed
previously in \refs{\EK, \Nakatsu, \TakSE, \Zabrodin}.)

  The integrability allows to find the unknown coefficients in the
Laurent expansion of the phase
\VPET. All the information about them is contained in the so called
$\tau$-function, which is the generating function for the vector
fields corresponding to the flows $H_n$. Its existence follows from
the zero-curvature condition \zerocur\ and can be proved in the
standard way \TakTak. The coefficients are related to the $\tau$-function as
follows
\eqn\vevv{ v_{n} = {\p\log\tau  \over \p t_{n}},
\qquad 
\phi(E ) = {i} \log { \tau(E+i/2R) \over \tau(E-i/2R) }. 
} The $\tau$-function can thus be considered as the generating
function for the perturbations by vertex operators. As we will show in
sec. 8, the logarithm of the $\tau$-function gives the free energy of
the Euclidean theory compactified at radius $R$.
 
 \subsec{The dispersionless (quasiclassical) limit}

Let us consider the quasiclassical limit $E \to -\infty$ when all
allowed energies are large and negative.  In this limit the integrable
structure described above reduces to the dispersionless Toda hierarchy
\refs{\Krichever, \TakTakb, \TakTak}, where the operators $\hat E$ and
$\oo$ can be considered as a pair of classical canonical variables
with Poisson bracket
\eqn\ccrom{ \{  \o, E  \} = \o. }
Similarly, all operators become $c$-functions of these variables.
The Lax operators can be identified, by  eq. \LaxPepm, 
with the phase space coordinates $\zpm$.
The functions $\xpm(\o, E) $  are given by eq.  
\eqn\zoom{
\zpm(\o, E)= e^{-{1\over 2R}\chi}\o^{\pm 1}
\left(1+\sum_{k\ge 1} a_{\pm k}(E)\ \o^{\mp k/R}\right),
} 
where we have introduced the ``string susceptibility"  
\eqn\suscep{ \chi   = -R \p_E\phi =  
\p_E^2\log\tau.}

 The equations \STREQ\ and \HamOp\ in the quasiclassical limit give
\eqn\cSEQ{
  \{ \zm,  \zp\}= {1},} 
\eqn\conteq{\zp\zm=M_+ = M_-.
} 
It follows from the first equation, \cSEQ, that eq. \zoom\ defines a
canonical transformation relating the phase-space coordinates $ \xp,
\xm$ to $\log \o$ and $E$. Moreover, the dressing procedure itself can
be interpreted as a canonical transformation between $ \xp, \xm$ and
the ``bare'' coordinates
\eqn\opmmm{ \o_+ =\sqrt{-E}\ \o, \ \ \  \o_-=\sqrt{-E}/\o.}
  The second equation, \conteq,  which is a deformation of the relation
\eqn\barevarO{
\o_+\o_-=-E}
satisfied by the ``bare" variables, yields the form of the perturbed
one-fermion trajectories. 

Let us consider the highest trajectory that defines the shape of the Fermi sea.
The variable $E$ then should be interpreted as the Fermi energy, which is
equal to (minus) the chemical potential,
$ E_F = -\mu.$
Taking into account the expression for $M_\pm$ through $\zpm$
given by the r.h.s. of \OrlSPepm, we can expand the right-hand side either in 
$\xp$ or in $\xm$
\eqn\TODApr{
\zp\zm=\sum _{k\ge 1}  k t_{\pm k} \  \zpm^{ k/R}  +\mu  +
\sum_{k\ge 1} v_{\pm k}\  \zpm^{-k/R}.}
(These expansions are of course convergent only for sufficiently large
values of $\xpm$.)  Eq. \TODApr , which is the dispersionless string
equation, is a particular case of the relation \TOD, which was derived
for an arbitrary potential.  In this way we found a direct geometrical
interpretation for the dispersionless string equation in terms of the
profile of the Fermi sea.

  A simple procedure to
calculate the coefficients $a_{\pm k}$  has been suggested in \IK .
First, note that if all $t_{\pm k}$ with $k>n$ vanish,
the sum in \zoom\ can be restricted to $k\le n$.
Then it is enough to substitute the expressions \zoom, with $E=-\mu$, in the
profile equations \TODApr\ and compare the coefficients in front of 
$\o^{\pm k/R}$. The result will give the generating function of one point
correlators, which is contained in the inverse function
$\o(\zpm)$. In the next section we will apply this
method  to calculate the profile of the Fermi sea in the 
important case of the Sine-Liouville string theory.

\newsec{ Solution for Sine-Gordon coupled to gravity}

 The simplest nontrivial string theory with time-dependent background
is the Sine-Gordon theory coupled to gravity known also as
Sine-Liouville theory. It is obtained by perturbing with the
lowest couplings $t_1$ and $t_{-1}$. It can be easily seen that in
this case the expansion \zoom\ consists of only two terms
 \eqn\zomSG{
z_\pm = e^{-{1\over 2R}\chi}  \o^{\pm 1} (1+ a_\pm \o ^{\mp {1\over R}} ).}
The  procedure described in \IK\  gives the following result
for the susceptibility $\chi$ and the coefficients $a_\pm$:
\eqn\coefzom{\eqalign{
&\mu e^{ {1\over R} \chi} - 
\left(1-{1\over R}\right)\tp\tm e^{{2R-1\over R^2} \chi} =1,
\cr
&a_\pm = \tmp \ e^{{R-1/2\over R^2} \chi}.  } }
The first equation in \coefzom\ for the susceptibility
$\chi=\p^2_\mu \log\tau$ was found in \KKK\foot{Since the papers
\refs{\KKK,\AK,\IK,\KZ} considered the case of vortex perturbations, the
duality transformation $R\to 1/R$ should be performed before 
comparing the formulas.} and it was shown to reproduce
the Moore's expansion of the free energy $F=\log\tau $ \MOORE.
The equation \zomSG\ with $a_\pm$ given in \coefzom\
was first found in \AK\ in the form
\eqn\CORKA{  e^{R h_\pm}- x_\pm e^{h_\pm}=1,\qquad x_\pm=e^{- \chi/2R^2} 
a_\pm \zpm^{-1/R},   }
where $h_\pm(\zpm)$ is the generating function of the one-point
correlators
\eqn\CORRE{h_\pm(\zpm)=\sum_{n=1}^\infty {1\over n}\zpm^{-n/R} {\p^2 F\over\p
\mu\p t_{\pm n}}. }
It is related to $\o(\zpm)$ as follows
\eqn\omh{\o(\zpm)=e^{{1\over 2R} \chi}\zpm e^{-R h_\pm(\zpm)}.
}

The equation \coefzom\ for the susceptibility imposes some
restrictions on the allowed values of the parameters $\tpm$ and $\mu$.
Let us consider the interval of the radii $\hf<R<1$, which is the most
interesting from the point of view of the conjectured duality between
the Sine-Liouville theory and the 2D string theory in
 the black hole background
at $R=2/3$ \FZZ.  In this interval, if we choose 
 $\tp\tm<0$, a real solution for $\chi$ exists only  for $\mu>0$.  
On the other hand, if $\tp\tm>0$, the allowed interval is
$\mu> \mu_c$, where the critical value is negative:
\eqn\crmu{
\mu_c=-\left(2-{1\over R}\right)\left({1\over R}-1\right)^{1\over 2R-1}
\left( \tp\tm \right)^{R\over 2R-1}.} 
 Let us note that  \crmu\ was
interpreted in \HSU\ as a critical point of the pure 2D gravity type.

The solution \zomSG-\coefzom\ allows us to find the explicit shape of
the Fermi sea in the phase space. In the model of the type I
considered so far there is an additional restriction for the
admissible values of parameters since the boundary of the Fermi sea
should be entirely in the positive quadrant $\zpm >0$.  This forces us
to take $\tpm>0$.  The general situation is shown on Fig. 1a.  The
unperturbed profile corresponds to the hyperbola asymptotically
approaching the $\zpm$ axis, whereas the perturbed curves deviate from
the axes by a power law. We see that there is a critical 
value of $\mu$, where the contour forms a spike.
It coincides with the critical point  $\mu =\mu_c$ given by \crmu.
At this point the quasiclassical description breaks down. 
On the Fig. 1b, the physical time evolution of a profile is demonstrated.

\figd{Profiles of the Fermi sea ($\xpm=x\pm p$) 
in the theory of the type I at $R=2/3$.  Fig. 1a contains several
profiles corresponding to $\tp=\tm=2$ and values of $\mu$ starting
from $\mu_c=-1$ with a step $40$.  For comparison, we also drew the
unperturbed profile for $\mu = 100$.  Fig. 1b shows three moments of
the time evolution of the critical profile at $\mu = -1$.}
{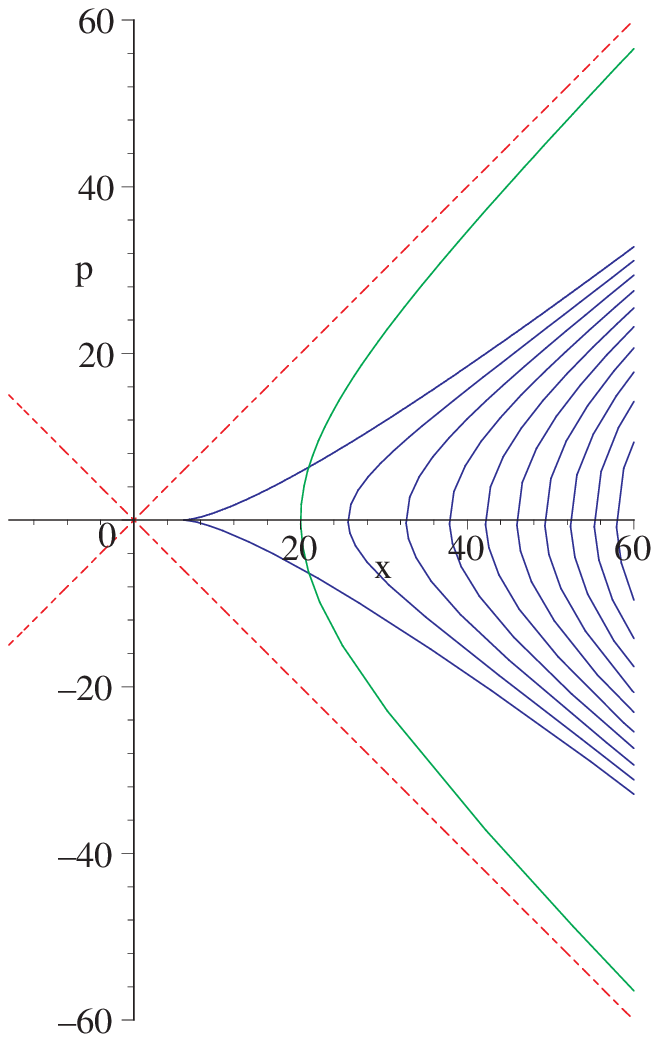}{4.5cm}{1cm}{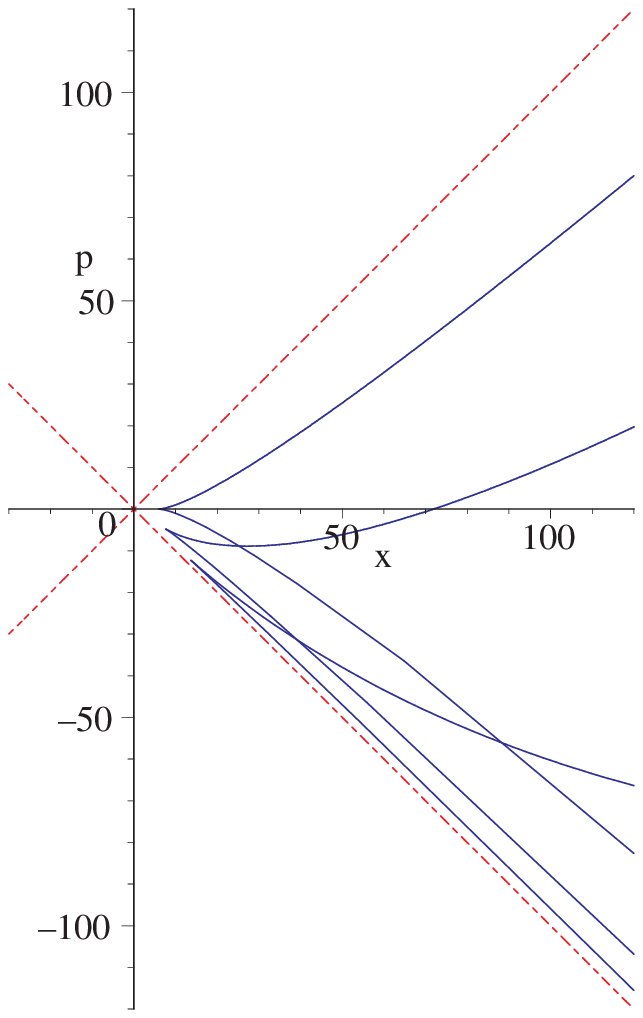}{4.5cm}

In the same way one can find the solution in the classical limit of
the theory of the type II described in Appendix A. In this case we can
introduce two pairs of perturbing potentials describing the
asymptotics of the wave functions at $\xpm \to\infty$ and $\xpm
\to-\infty$. For sufficiently large $\m$ the Fermi sea consists of two
connected components and the theory decomposes into two theories of
type I. However, in contrast to the previous case, there are no
restrictions on the signs of the coupling constants. When $\mu$
decreases, the two Fermi seas merge together at some critical value
$\m^*$. This leads to interesting (for example, from the point of view
of the Hall effect) phenomena, which we intend to discuss
elsewhere. Here we will only mention that, depending on the choice of
couplings, it can happen that for some interval of $\mu$ around the
point $\mu^*$, the Toda description is not applicable.

\newsec{Free energy of the perturbed background}

As we mentioned, the perturbations considered above appear in a theory
at the finite temperature $T=1/\b$, $\b= 2\pi R$.  The free energy per
unit volume $\CF$ is related to the partition function $\CZ$ by
\eqn\FRZ{\CZ =
e^{-\b\CF}.  }
 The free energy is considered as a function of the
chemical potential $\mu$, so that the number of fermions is given by
\eqn\dFdm{
  N = \p \CF/\p \mu.}

We have seen that any perturbation can be
characterized by the profile of the Fermi sea.
For a generic profile, the number of fermions is given by 
the volume of the domain in the phase space occupied by
the Fermi liquid
\eqn\Ndom{
  N= \dpi \mathop{\int\int}\limits_{\fs} d\xp d\xm .  } It is implied
that the integral is regularized by introducing a cut-off at distance
$\sqrt{\Lambda}$. For the unperturbed ground state \TRAJ, one
reproduces from \dFdm\ and \Ndom\ the well known result for the
(universal part of the) free energy
\eqn\FEN{ \CF  =  {1\over 4\pi} \mu^{2}\log\mu.}

 The authors of \DMP\ identify the partition function \FRZ\ for the
tachyon backgrounds studied in sect. 6 with the $\tau$-function of the
Toda hierarchy, defined by \vevv.  In this section we will reproduce
this statement by direct integration over the Fermi sea, adjusting the
above derivation of \FEN\ to the case of a general profile of the form
\TODApr.

 Since the change of variables described
by eq. \zoom\ is canonical (see eq. \cSEQ), one can  rewrite 
the integral \Ndom\ as 
\eqn\avV{
N={1\over 2\pi}\int\limits^{-\mu } dE \int\limits_{\bom(E)}^{\bop(E)}
{d\o \over \o} .  }
 The limits of integration over $\o$ are determined by the
cut-off.\foot{The integral over $E$ is also bounded from below, but we
do not need to specify the boundary explicitly.}  If we put it at the
distance $\zpm=\sqrt{\Lambda}$, they can be found from the equations
\eqn\grcon{
\zpm(\bopm(E),E)=\sqrt{\Lambda}.
}
 Taking the derivative with respect to $\mu$, we obtain
\eqn\numb{
\p_{\mu}N=-\dpi \log{\bop(-\mu)\over \bom(-\mu)}.}
It is enough to keep only the leading order in the cut-off $\Lambda$
for the boundary values $\bopm$.  From \grcon\ and \zoom\ we find to
this order
\eqn\grsolf{
\bopm=\left(\Lambda e^{\chi/R}\right)^{\pm 1/2}.
}
Combining \numb\ and \grsolf, we find
\eqn\frener{
\p_{\mu}N=
-\dpi \log\Lambda- {1 \over 2\pi R} \chi .
}
Taking into account the relation \dFdm\ between the free energy and
$N$, we see that 
\eqn\frenercom{
\CF=- {1\over\beta }\FT,  \qquad \beta =2\pi R.  }  
  As a result, if
we compactify the theory at the time circle of the length $\beta$, our
free energy coincides with the logarithm of the Toda
$\tau$-function.   

The  perturbed  flow of the Fermi liquid is non-stationary with respect to the
physical time $t$ associated with the original Hamiltonian $H_0$. 
In sec. 5 we constructed a new Hamiltonian, which preserves the shape of
the Fermi sea. The latter coincides with the phase-space trajectory 
\eqper\ at $E=-\mu$.  
Using the fact that $\p_{\mu}H(\zp,\zm)=0$, one can check that
\eqn\FH{
\CF=\langle H(\zp,\zm)+\mu\rangle.
} 
Note that the function $H(\xp,\xm)$ is obtained by solving the profile equations 
\TODApr\ with respect to $\m$.
One can see, looking at the large $\xpm$ asymptotics of the 
profile equations, that it  can be written in the
following compact form:
\eqn\DVHam{
H(\zp,\zm)=H_0+\sum\limits_{k\ge 1} k\ ( t_k H_k+t_{-k} H_{-k}).}  Here
the Hamiltonians $H_n(\xp,\xm;  \vec t)$ are given by equations \hashka, 
with $\o$ and $E$ expressed as functions of $\xp$ and $\xm$ through
eq. \zoom.

\newsec{ Conclusions }

In this paper we studied the $2D$ string theory in the presence of
arbitrarily strong tachyonic perturbations. In the matrix model
language, we dealt with the singlet sector of the MQM described by a
system of free fermions in inverted quadratic potential.  In the
quasiclassical limit the state of the system is described by the shape
of the domain in the phase space occupied by the fermionic liquid (the
profile of the Fermi sea). In this limit we formulated functional
equations, which determine the shape of the Fermi sea, given its
asymptotics at two infinities.  There is some analogy between our
problem and the problem of conformal maps studied in
\refs{\wz,\kkvwz,\bmrwz,\mwz}.

This allows to calculate various
physical quantities related to the perturbed shape, such as the free
energy of the compactified Euclidean theory, in terms of the
parameters defining the asymptotic shapes.

 For a particular case of perturbations generated by vertex operators
(in- and out-going with discrete equally spaced imaginary energies),
the system solves the Toda lattice hierarchy where the Toda ``times''
correspond to the couplings characterizing the asymptotics of the
profile of the Fermi sea.  Using the ``light cone'' representation in
the phase space, we developed the Lax formalism of the constrained
Toda hierarchy, where the constraint coincides with the string
equation found recently in \IK. In the dispersionless limit, the Lax
formalism reproduces the functional equations for the shape of the
perturbed Fermi sea and allows to give a direct physical meaning to
the Toda mathematical quantities.  In particular, we identified the
two spectral parameters of the Toda system with the two chiral
coordinates $\xpm$ in the target phase space.

Let us list the other possible applications of our formalism:

\noindent
$\bullet$ Calculation of the tachyon scattering matrix for 2D string
theory in nontrivial backgrounds.

\noindent
$\bullet$ Analysis of the systems with more complicated than
Sine-Liouville backgrounds. In particular, the quasiclassical analysis
of the sections 4 and 7 are not limited to any lattice of tachyon
charges and can be used for a mixture of any commensurate or
non-commensurate charges. It provides a way of studying the whole space
of possible tachyonic backgrounds of the $2D$ string theory.
 
\noindent
$\bullet$ Investigation of the thermodynamics of the $2D$ string
theory in Sine-Liouville type tachyonic backgrounds characterized by a
lattice of discrete Matsubara energies \KZ, \AKKII. This is supposed
to be the thermodynamics of the dilatonic black hole, according to the
conjecture of \FZZ.
  
\noindent
$\bullet$ The type 2 theory can be used to model the situation where
either two quantum Hall droplets  approach each other
or one droplet is about to split in two.

\newsec{ Problems and proposals  }
 
We list also some unsolved problems, which could
be approached by our formalism:

\noindent
$\bullet$  To find the metric of the target space for the Sine-Liouville 
string theory.
 
 \noindent
$\bullet$  To construct  the discrete states of the 2D string theory (at any
rational compactification radius) in the framework of our formalism.

\noindent
$\bullet$ A few important unsolved problems concern the relation between the
MQM and CFT formulations of the 2D string theory. We still don't have
a precise mapping of the vertex and vortex operators of the MQM to the
corresponding operators of CFT, in spite of the useful suggestions of
the papers \MSS, \POLCHINSKI. As the result, we did not manage to
match the Sine-Liouville/Black Hole correlators calculated by \FZZ,
\TECHNER, and \JAP\ from the CFT approach with the corresponding
correlators found in \AK\ from the Toda approach to the MQM. Another
related question is how can the target space integrable structure  be seen in the  CFT formulation of 2D string theory.

\noindent
$\bullet$   Description and quantitative analysis of the mixed (vertex\&vortex)
perturbations in the 2D string theory by means of the MQM. 
We hope  that the ``light-cone'' formalism might  
be appropriate to study such perturbations,
in spite of the absence of integrability.\foot{ An interesting proposal to describe nontrivial string
backgrounds on the CFT side was given long ago in \WITTENGR\ and
elaborated in \MUKHIVAFA,\GOSHALVAFA. It suggested to parameterize the
moduli space of the 2D string theory perturbed in both sectors, by a
conifold. This structure may be hidden in the 3 matrix model proposed
below.}

Finally, keeping in mind this interesting problem, let us propose here
a 3-matrix model whose grand canonical partition function $ \CZ(\mu,
t_n,\tilde t_n) = \sum_{N=0}^\infty e^{-2\pi R\mu N} \CZ_N( t_n,
\tilde t_n)$ is the generating function of correlators for both types
of perturbations: the parameters $\tilde t_n$ are the couplings of the
vortex operators, in the same way as $t_n$'s are the couplings of the
tachyon (vertex) operators throughout this paper. As usual, the
chemical potential $\mu$ plays the role of the string coupling. The
partition function at fixed $N$ is represented by the integral over
two hermitian matrices $\Xpm$ and one unitary matrix, the holonomy
factor $\Omega$ around the circle
\eqn\partvor{
\CZ_N(t_n, \tilde t_n)=
 \int  [\CD \Xp]_{ \{ t_n\}}
[\CD \Xm ]_{\{ t_{-n}\} }
  [d\Omega ]_{ \{\tilde t_{\pm n}\}}\ 
\ e^{  \Tr \left(\Xm\Xp -  q \Xm\Omega \Xp  \Omega^{-1} \right)},
}
where we denoted $q= e^{i2\pi R}$ and introduced the following matrix integration measures   
\eqn\deforpm{
  [d \Xpm]_{ \{ t_{\pm n}\}}= d \Xpm \ 
e^{ R \sum_{n>0}  t_{\pm n}\Tr   \Xpm^{ n/R } }, }
\eqn\actvervor{   [d\Omega ]_{\{ \tilde t_{\pm n}\}}= [d\Omega ]_{_{SU(N)}}
e^{\sum _{n\ne 0} \tilde t_n \Tr \Omega^{n} } .}
This 3-matrix integral is nothing but the Euclidean partition function
of the upside-down matrix oscillator on a circle of length $2\pi R$
with twisted periodic boundary conditions $\Xpm(\b)=\Omega^\dagger
\Xpm(0) \Omega$, needed for the introduction  of the winding modes 
(vortices) in the MQM model of the 2D string theory \BULKA.  It is
trivial to show that in particular case of all $t_n=0$ this 3-matrix
model reduces to the model with only vortex perturbations of the paper
\KKK. It can be also shown \AKKII\ that in the case of all $\tilde
t_n=0$ it reduces to the Euclidean version of the model with only
tachyonic perturbations studied in this paper.  The crucial point here
is that the sources for tachyon perturbations are introduced in a
single point on the time circle, similarly to the perturbation
imposing the initial conditions on the wave function \asswave.
The model \partvor\ reduces the problem of the study of the backgrounds
of the compactified 2D string theory with arbitrary tachyon and
winding sources, to a three-matrix integral.

We learned from R. Dijkgraaf and C. Vafa that they discovered a
similar Toda structure in the $c=1$ type theory arising in connection
to the topological strings.

\bigskip
\noindent{\bf Acknowledgements:} We would like to thank A. Boyarsky,
G. Moore, A. Sorin, C. Vafa, and especially P. Wiegmann for useful
discussions.  Two of the authors (V.K. and I.K.) thank the string
theory group of Rutgers University, where a part of the work was done,
for the kind hospitality. This work of S.A. and V.K. was partially
supported by European Union under the RTN contracts HPRN-CT-2000-00122
and -00131. The work of S.A. and I.K. was supported  in part by
European network EUROGRID HPRN-CT-1999-00161.


\appendix{A}{Theory of type II}

In the theory of type II, the fermions are defined on the whole real line.
In this case one should introduce two sets of functions
describing, in the quasiclassical limit, fermions at different sides of the potential.
They are defined on right (left) semi-axis by
\eqn\waveff{\eqalign{
\prepm(\xpm)& = {1\over\sqrt{2\pi}}  
{\xpm^{\pm iE-\hf}\over \sqrt{1+e^{2\pi E}}}\qquad \ \ \ (\xpm>0),\cr
\plepm(\xpm)& = {1\over\sqrt{2\pi}} 
{(-\xpm)^{\pm iE-\hf}\over \sqrt{1+e^{2\pi E}}}\qquad(\xpm<0)
}}
and the continuation to the other semi-axis is performed according to
the  rule 
$\xpm \rightarrow e^{\mp \pi i}\xpm$. This gives, for $\xpm>0$
\eqn\waveanff{\eqalign{
\prepm(-\xpm)& =\pm i e^{\pi E}\prepm(\xpm), \cr
\plepm(\xpm)& =\pm i e^{\pi E}\plepm(-\xpm) .
}}
The functions \waveff\ satisfy the orthonormality and completeness
conditions
\eqn\normlr{
\langle \piepm{a}|\pieepm{b}\rangle=\delta^{ab}\delta(E - E'), }
\eqn\compllr{\int_{-\infty}^{\infty}dE\, \sum_{a} 
\overline{\piepm{a}(\xpm)}\piepm{a}(\xpm') 
= \delta( \xpm-\xpm'), } where the indices $a,b=(>,<)$ label the right
and left wave functions and the scalar product is defined as in
\normpm\ but with the integral over the whole real axis.
 
The unitary operator relating $\xpm$
representations is given by the Fourier kernel on the whole line
$K(\xm,\xp) = {1\over \sqrt{2\pi}} e^{ i \xp\xm}$
\eqn\Fourie{[\hat S \psip](\xm)= 
 \int_{-\infty}^{\infty}
d\xp \, K(\xm,\xp)  \psip(\xp).}
It acts on the eigenfunctions \waveff\ by the following matrix
\eqn\Smatr{
\left[\hat S^{\pm 1}  \piepm{a} \right](\xmp)   
=\sum_b [S^{\pm 1}(E)]^{ab} \piepm{b}(\xmp), \qquad S(E)
=\CR(E)\pmatrix{1 & - i e^{\pi E} \cr
-i e^{\pi E} & 1},
}
where 
\eqn\rfacts{
\CR(E)  = {1\over \sqrt{2\pi}}
e^{-{\pi\over 2} (E- i/2)}  \Gamma( iE + 1/2). 
} 
In this case the reflection coefficient $\CR(E)$ is not a pure phase 
because of the tunneling through the potential described by the off-diagonal elements of the $S$-matrix.
Nevertheless, the whole $S$-matrix is unitary
\eqn\unitS{S^{\dagger}(E)S(E)=1.
}
The scalar
product between left and right states is given by
\eqn\scprlr{
\langle\psim|K| \psip\rangle=  
\int_{-\infty}^\infty d\xp d\xm\,\overline{\psim(\xm)}K (\xm,\xp)
\psip(\xp).} 
It is clear that the matrix of scalar products of states \waveff\
coincides with $S(E)\delta(E-E')$.
The corresponding completeness condition is
\eqn\comppmrl{
\int_{-\infty}^{\infty}dE\, \sum_{a,b} 
\overline{\piem{a}(\xm)} \left(S^{-1}\right)_{ab}\piep{b}(\xp)  
= {1\over \sqrt{2\pi}} e^{-i\xp\xm}.
}

\listrefs
 
\bye